# Event by Event analysis of Maximum Pseudo-rapidity Gap Fluctuation in High energy Nucleus-nucleus Collisions

Swarnapratim Bhattacharyya[1], Maria Haiduc[2], Alina Tania Neagu[2] and Elena Firu[2]

[1]Department of Physics, New Alipore College, L Block, New Alipore, Kolkata 700053, India

Email: swarna_pratim@yahoo.com

[2]Institute of Space Science, Bucharest, Romania

## Abstract

A detailed study of event-by-event maximum pseudo-rapidity gap fluctuations in relativistic heavy-ion collisions in terms of the scaled variance $\omega$ has been carried out for $^{16}$O-AgBr, $^{28}$Si-AgBr and $^{32}$S-AgBr interactions at an incident momentum of 4.5 AGeV/c applying the multiplicity cut ($N_s > 10$). For all the interactions the values of scaled variance are found to be greater than zero indicating the presence of strong fluctuation of maximum rap gap values in the multiparticle production process.  The event by event fluctuations are found to decrease with the increase of average multiplicity of the interactions.  Experimental analysis has been compared with the results obtained from the analysis of events simulated by generating random numbers (MC-RAND events) and also by Ultra Relativistic Quantum Molecular Dynamics (UrQMD) model. UrQMD model simulated values of event by event fluctuations of maximum rapidity gap are less than the corresponding experimental values.

# Introduction

High energy nucleus-nucleus collisions deals with the production of copious amounts of various particles and multiplicity of such particles are extremely important to investigate the detailed characteristics of particle production process. The goal of the current heavy-ion collision physics is to study the properties of the phase transition between the quark–gluon plasma (QGP) phase and the ordinary hadronic phase. Fluctuation measurements in heavy-ion collisions have a good chance of being the signals of the QGP formation. The study and analysis of fluctuations are an essential method to characterize a physical system. Fluctuations of different observables of the final-state particles act as a probe to study the phase transition [1-4]. Fluctuation in particle multiplicity and momentum distribution has been found to be very useful to explore the thermalization and statistical behaviour of the produced particles [5-9]. Multiplicity fluctuation can tell us whether a global thermalization has been achieved [10]. In nucleus-nucleus collisions, transverse energy is an extensive global variable. Transverse energy is also an indicator of the energy density achieved in the collisions. Studies of transverse energy fluctuations are very important [11-12] as energy density is directly related to the formation of QGP. Charge fluctuations [13-15] are sensitive to the unit charge of the underlying system. Quarks have fractional electric and baryonic charges. Therefore, the fluctuations of those charges in a QGP and a hadronic matter are clearly distinguishable. All these are interesting and deserve careful study. There has been evidence for occurrence of multiplicity fluctuations in the pseudo-rapidity distributions [16]. The existence of such fluctuations would give information on the substructure in space-time of collision region. Large fluctuations in pseudo-rapidity window have been observed in cosmic ray events and in hadron-hadron, nucleus-nucleus and hadron-nucleus interactions at accelerator energies [17-18]. They have been interpreted in terms of several models: as a possible indication of hadronic phase transition, as Cherenkov radiation or simply originating from a cascade mechanism. The rapid development in the field of pion multiplicity fluctuations in recent years is related to the large amount of high multiplicity data from well-known heavy ion experiments at CERN-SPS and BNL-RHIC. Although some progress has been made in understanding the fluctuation phenomena, a lot of questions remain unanswered.

In the recent years studies of event by event fluctuations have gained immense popularity among the scientists. The study of event-by-event fluctuations in high-energy heavy-ion collisions may provide us more information about the multiparticle production dynamics [19-22]. Event-by-event analysis is potentially a powerful technique to study relativistic heavy-ion collisions, as the magnitude of fluctuations of various quantities around their mean values is controlled by the dynamics of the system. It is believed that a detailed study of each event produced in high-energy nucleus-nucleus collision, may reveal new phenomena occurred in some rare events for which favourable conditions may have been created. High-energy nucleus-nucleus collisions produce a large number of particles. The study of a single event with large statistics can reveal very different physics than the analysis of averages over a large statistical ensemble. Event-by-event fluctuations may provide us information about the heat capacity [21, 23-25], possible equilibration of the system [26-34] or about the phase transition [25, 35]. An important characteristic of muliparticle production that deserves special attention is the maximum pseudo-rapidity gap. Maximum pseudo-rapidity gap ($\Delta\eta_{max}$) in an event has been defined as the difference of maximum and minimum pseudo-rapidity values of the two produced shower particles in an event. From the maximum pseudo-rapidity gap values one can get information for the particle production process. Diffractive processes are expected to contribute mainly due to the larger values of pseudo-rapidity gaps [36]. Maximum pseudo-rapidity gap values are important to study long range correlations suppressing the contributions from resonance and mini-jets [37].

So far investigations of fluctuations of rapidity gaps have been carried out by different groups [38-41] to extract the dynamical signal of particle production process. None so far have studied the event by event fluctuations of maximum pseudo-rapidity gap ($\Delta\eta_{max}$). This study may also be a potential source of information towards the hidden dynamics of particle production process.

Our aim in this paper is to carry out a detailed analysis of the event-by-event fluctuations of maximum pseudo-rapidity gap in terms of the scaled variance $\omega$ for $^{16}$O-AgBr, $^{28}$Si-AgBr and $^{32}$S-AgBr interactions at an incident momentum of 4.5 AGeV/c. Experimental analysis has been compared with the results obtained from the analysis of events simulated by generating random numbers (MC-RAND events) and also by Ultra Relativistic Quantum Molecular Dynamics (UrQMD) model.

## Experimental Details

The present analysis is based on the interactions of $^{16}$O, $^{28}$Si and $^{32}$S projectiles at an incident momentum of 4.5 AGeV/c with AgBr as target present in nuclear emulsion. The data were obtained by exposing the stacks of NIKFI-BR2 emulsion pellicles of dimension 20cm×10 cm ×600$\mu$m horizontally to 4.5 AGeV/c $^{16}$O, $^{28}$Si and $^{32}$S projectile beams at Dubna Synchrophasotron. In order to get the required data for the present analysis NIKFI-BR2 emulsion pellicles of dimension 20cm×10 cm ×600$\mu$m were irradiated by the $^{16}$O, $^{28}$Si and $^{32}$S sbeam at 4.5 AGeV/c obtained from the Synchrophasatron at the Joint Institute of Nuclear Research (JINR), Dubna, Russia [42].

The scanning of the emulsion plates has been carried out as a part of a planned programme in which events of certain types can be sought in a systematic way. We have applied the double scanning method along the track of the incident beam starting from the entry point of the beam into the emulsion plate fast in the forward direction and slowly in the backward direction to select the primary interactions. Scanning has been performed under 100× magnification. Slow scanning in the backward direction ensures that the events chosen did not include interactions from the secondary tracks of the other interactions. During scanning, we have excluded the events those have scattered at small angles without showing any indication of excitation or break up of the target nucleus satisfying the criteria of elastic collisions. Events produced by electromagnetic dissociation of the projectiles have also been excluded from our consideration. Excluding elastic interactions and electromagnetic interactions, we have considered only the events due to inelastic collisions. After scanning those events were selected where the incident beam lie within $3^0$ from direction of the main beam in the pellicle. This criterion ascertains that the real projectile beam has been selected for the analysis. Events showing interactions close to the emulsion surface and glass surface (interactions within 20 $\mu$m from the top and bottom surface of the pellicle) were excluded from our consideration. Rejection of such events reduces the losses of tracks and minimizes the uncertainties in the measurements of emission and azimuthal angles.

According to the terminology of nuclear emulsion, particles emitted from an interaction (called an event or a star) are classified into four categories, namely the shower particles, the grey particles, the black particles and the projectile fragments [43]. Characteristics of these particles are given below.

**Shower particles**: The tracks of particles having ionization I less or equal to $1.4I_0$ are called shower tracks. $I_0$ is the minimum ionization of a singly charged particle. The shower particles are mostly pions (about more than 90%) with a small admixture of kaons and hyperons (less than 10%). These shower particles are produced in a forward cone. The velocities of these particles are greater than $0.7c$ where $c$ is the velocity of light in free space. Because of such a high velocity, these particles are not generally confined within the emulsion pellicle. Energies of these shower particles lie in the GeV range.

**Grey particles**: Grey particles are mainly fast target recoil protons with energies up to 400 MeV. They have ionization $1.4\ I_0 \leq I < 10\ I_0$. Ranges of these particles are greater than 3 mm in the emulsion medium. These grey particles have the velocities lying between $0.3c$ and $0.7c$.

**Black particles**: Black particles consist of both singly and multiply charged fragments. They are fragments of various elements like carbon, lithium, beryllium etc with ionization greater or equal to $10I_0$. These black particles have the maximum ionizing power. They are less energetic and consequently they are short ranged. In the emulsion medium, ranges of black particles are less than 3 mm. The velocities of the black particles are less than $0.3c$. In emulsion experiments, it is very difficult to measure the charges of the target fragments. Therefore, it is not possible to identify the exact nucleus.

**Projectile Fragments**: The projectile fragments are the spectator parts of the incident projectile nucleus that do not directly participate in the interaction. They are emitted within the critical angle $\theta_c < 3°$. In terms of the incident projectile momentum per nucleon ($P_{inc}$) measured in GeV/c and the Fermi momentum ($P_{Fermi}$) of the nucleons of the projectile in GeV/c, the critical angle $\theta_c$ can be defined as $\theta_c = \frac{P_{Fermi}}{P_{inc}}$. The value of $P_{Fermi}$ can be calculated on the basis of Fermi gas model of the nucleus. Projectile fragments exhibit uniform ionization over a long range and suffer negligible scattering as they have almost the same energy or momentum per nucleon as the incident projectile. Here it may be mentioned that the projectile fragments are important to determine the centrality of collisions.

Nuclear emulsion medium consists of variety of nuclei like H, C, N, O, Ag and Br. In emulsion experiment, it is very difficult to measure the charges of the fragments emitted from the target and hence exact identification of the target is not possible. However, we can divide the major constituent elements present in the emulsion into three broad target groups namely hydrogen (H), light nuclei (CNO) and heavy nuclei (AgBr) on the basis of the number of heavy tracks ($N_h$) .The black and the grey tracks are collectively called the heavily ionizing tracks and the number of such tracks in an event is represented by $N_h$. Number of heavy tracks in an interaction is an important parameter and it helps in separating the events occurring due to the collisions of the projectile beam with the different types of target group present in emulsion medium. Usually events with $N_h \leq 1$ have occurred because of the collision between hydrogen nucleus and the projectile beam . This is because, when hydrogen nucleus is the target, on fragmentation the only proton will be emitted.  Events with $2 \leq N_h \leq 8$ are due to collisions of the projectile beam with the light nuclei (CNO). In case of interactions of the projectile beam with the light nuclei (CNO), it is evident that maximum number of heavy tracks cannot exceed eight. This corresponds to the largest charge of the light nucleus-the oxygen nucleus.  Events with $N_h > 8$ are due to collisions of the projectile beam with the heavy target (AgBr). It can easily be understood that when $N_h > 8$, it can be ascertained that the projectile has collided with such a target whose atomic number is greater than eight. It is clear that in this case, the target cannot be of the light target group (CNO). It will be the heavy target group (AgBr). In this way by counting the number of heavy tracks, one can determine the target group.

Now one point should be mentioned. The method of target identification as discussed above has some limitations in identifying the CNO target. But the method is quite accurate in selecting the interactions of the projectile beam with the AgBr target. We have earlier said those events with $2 \leq N_h \leq 8$ are due to collisions of the projectile beam with the light nuclei (CNO). However, when $2 \leq N_h \leq 8$, there is also a possibility of events generated from peripheral collisions between the projectile and AgBr target. Therefore,  there is a possibility that in this case, the events could be an admixture of CNO events and peripheral AgBr events.

Applying the method of target identification based on the number of heavy tracks total number of inelastic interactions can be divided into three target groups H, CNO and AgBr. For the present analysis we have not considered the events which are found to occur due to collisions of the projectile beam with H and CNO target present in nuclear emulsion. Our analysis has been carried out for the interactions with the AgBr target only. For the present study, we have selected 1057 events of $^{16}$O-AgBr, 514 events of $^{28}$Si-AgBr and 434 events of $^{32}$S-AgBr interactions. Average multiplicities of shower tracks of each interaction have been calculated and presented in table 1.

The emission angle ($\theta$) was measured for each track with respect to the beam directions by taking readings of the coordinates of the interaction point ($X_0$, $Y_0$, $Z_0$), coordinates ($X_1$, $Y_1$, $Z_1$) at the end of the each secondary track and coordinates ($X_i$, $Y_i$, $Z_i$) of a point on the incident beam. In case of shower particle multiplicity distribution the phase space variable used is pseudo-rapidity $\eta$. The relation $\eta = -\ln\tan\frac{\theta}{2}$ relates the variable $\eta$ with the emission angle $\theta$. In an emulsion experiment, the pseudo-rapidity is a convenient choice for the basic variable in terms of which the particle emission data can be analyzed. It is an approximation of the dimensionless boost parameter rapidity (additive under Lorentz boost) of a particle.

## Analysis and Results

As we are dealing with interactions at 4.5AGeV/c, problems may be caused by lower multiplicity events. To avoid this for the present analysis we have selected those events which have the number of shower particles ($N_s$) greater than 10. Average multiplicity of those events having number of shower tracks greater than 10 ($N_s > 10$) have been presented in table 2.

In order to calculate the event by event fluctuations of the maximum pseudo-rapidity gap ($\Delta\eta_{max}$) in $^{16}$O-AgBr, $^{28}$Si-AgBr and $^{32}$S-AgBr interactions at 4.5 AGeV/c, we have calculated the maximum pseudo-rapidity gap for the each event of each interactions. The quantification of event by event fluctuations of the maximum pseudo-rapidity gap has been performed with a variable $\omega$, called the scaled variance such that $\omega = \frac{\langle \Delta\eta_{max}^2 \rangle - \langle \Delta\eta_{max} \rangle^2}{\langle \Delta\eta_{max} \rangle}$. Event by event fluctuation of maximum rapidity gap signifies the correlated production of shower particles.

From the values of the maximum pseudo-rapidity gap ($\Delta\eta_{max}$) we have calculated the value of the scaled variance $\omega$ and presented the calculated values in table 3. From table 3 it may be pointed out that the values of scaled variance indicating event-by-event fluctuations of maximum rapidity gap are found to be greater than zero indicating the presence of strong fluctuation of maximum rap gap values in the multiparticle production process.

From the table 2 and table 3 it can easily seen that values of the event by event fluctuations of the maximum pseudo-rapidity gap decreases with the increase of average multiplicity <$N_s$> of the interactions. The variation of the variable $\omega$ with the average multiplicity of interactions ($<N_s>$) has been presented in figure1 for all the interactions. Error bar shown in the figures are statistical errors only.

In order to ascertain that the observed event by event fluctuations for maximum pseudo-rapidity gap are not due to mere statistics only, the comparison of the experimental results with the analysis of the event samples simulated by generating random numbers (MC-RAND) is extremely important. For the analyses of the experimental data, the events generated by the Monte Carlo simulation (MC-RAND events) can provide a kind of reference data sample in which dynamical correlations among the particles are completely destroyed. We have utilized the framework of independent emission hypothesis based on the following assumptions in order to generate the MC-RAND events as did in our previous publication [44].

(i) The particles have been produced independently of each other.

(ii) The multiplicity distributions of the Monte Carlo events should replicate the multiplicity distribution of the experimental data.

(iii) The pseudo-rapidity distribution of the simulated events should reproduce the pseudo-rapidity distribution of the experimental data.

To generate the MC-RAND events, the multiplicity and pseudo-rapidity distribution of the experimental data has been given as the input. The pseudo-rapidity values of the simulated events have been generated by generating random numbers. The MC-RAND events have the identical pseudo-rapidity and multiplicity distribution as the experimental data [47]. Consequently, the average multiplicities of the shower tracks for the MC-RAND events will be equal to those of the experimental events. We have generated Monte Carlo-simulated (MC-RAND) events for $^{16}$O-AgBr, $^{28}$Si-AgBr and $^{32}$S-AgBr interactions in the pseudo-rapidity phase space. Table 1 and table 2 represent the average multiplicity of all the interactions for the MC-RAND events without and with the multiplicity cut ($N_s > 10$) respectively. As before we have calculated the event by event fluctuation variable and presented the values in table 3 along with the experimental values for the events $N_s > 10$ for all the interactions. Comparing the values of the event by event fluctuations of experimental and randomised data sample from table 3 it may be seen that the values of the event by event fluctuations for the randomized data sample are significantly less than their experimental counterparts. This observation signifies that the experimental data reveals the true dynamical signal. No clear dependence of event by event fluctuations of maximum pseudo-rapidity gap on the average multiplicity of the interaction has been observed. Figure 1 depicts the variation of the variable $\omega$ with the average multiplicity of interactions for the MC-RAND events also.

The experimental results have also been compared with those obtained by analyzing events generated by the Ultra Relativistic Quantum Molecular Dynamics (UrQMD) model. UrQMD model is a microscopic transport theory, based on the covariant propagation of all the hadrons on the classical trajectories in combination with stochastic binary scattering, color string formation and resonance decay. It represents a Monte Carlo solution of a large set of coupled partial integro-differential equations for the time evolution of various phase space densities. The main ingredients of the model are the cross sections of binary reactions, the two-body potentials and decay widths of resonances. The UrQMD collision term contains 55 different baryon species (including nucleon, delta and hyperon resonances with masses up to 2.25 GeV/c$^2$) and 32 different meson species (including strange meson resonances), which are supplemented by their corresponding anti-particle and all isospin-projected states. The states can either be produced in string decays, s-channel collisions or resonance decays. This model can be used in the entire available range of energies from the Bevalac region to RHIC. For more details about this model, readers are requested to consult [45-46]. We have generated a large sample of events using the UrQMD code (UrQMD 3.3p1) for $^{16}$O-AgBr, $^{28}$Si-AgBr and $^{32}$S-AgBr interactions in the pseudo-rapidity phase space [47]. We have also calculated the average multiplicities of the shower tracks for all the three interactions in case of the UrQMD data sample. Average multiplicities of the shower tracks in case of UrQMD data sample have been presented in table 1 along with the average multiplicity values of shower particles in the case of the experimental and MC-RAND events. The analysis of event by event maximum rap gap fluctuations has been repeated with the UrQMD simulated events with the multiplicity cut ($N_s > 10$). Average multiplicity of all the interactions for the UrQMD events after applying the multiplicity cut ($N_s > 10$) have been presented in table 2. The calculated values of scaled variance signifying event by event fluctuations for the UrQMD simulated events have been presented in table 3 for $^{16}$O-AgBr, $^{28}$Si-AgBr, $^{32}$S-AgBr interactions. From the table 3 it may be seen that the UrQMD model simulated values of event by event fluctuations of maximum rapidity gap are less than the corresponding experimental values. So UrQMD model reflects weaker dynamical fluctuations in comparison to the experimental values. From the table it can also be seen that for the UrQMD events the event by event fluctuations of the maximum pseudo-rapidity gap decreases with the increase of average multiplicity of the interactions. The variation of the variable $\omega$ with the average multiplicity of the interactions has been presented in figure 1 in case of the UrQMD simulated events along with the experimental and MC-RAND events.

**Discussion of Systematic Errors**

Before going to the conclusion of our analysis, it would be relevant to present a discussion on systematic errors of the data. Detailed analysis on systematic errors for the same data set has been presented in our earlier publication [48-49]. From our previous papers [48-49], we find that the efficiency of identifying the shower particles contribute to the systematic error of the analysis. We have shown that [48-49] applying the "along the track" scanning procedure with the help of two independent observers helps us to increase the scanning efficiency to more than 99% and hence to minimize the systematic errors to less than 1%. It has been mentioned in [48-49] that "along the track" scanning method gives reliable event samples because of its high detection efficiency. In nuclear emulsion detector tracks of $e^{+}e^{-}$ pair originating from γ –conversion may cause problem in the identification of pions [48-49]. These tracks are produced at a distance from the interaction point after travelling through certain radiation lengths. In order to exclude such tracks lying close to shower tracks near vertex, special care was taken while performing angular measurements. Usually all the shower tracks in the forward direction were followed more than 100–200 $\mu$m from the interaction vertex for angular measurement. The tracks due to $e^{+}e^{-}$ pair can be easily recognized from the grain density measurement, which is initially much larger than the grain density of a single shower track. It may also be mentioned that the tracks of an electron and positron when followed downstream in nuclear emulsion showed considerable amount of Coulomb scattering as compared to the energetic charged pions. Such $e^{+}e^{-}$ pairs were eliminated from the data and consequently they contribute almost nothing (~0.01%) to the systematic errors [44, 48-49].

It has been mentioned earlier that the shower particles are mostly pions (more than 90%) with a small proportion (less than 10%) of kaons and hyperons among them. The presence of K-mesons, hyperons and any other mesons among the pions are treated as contaminations. The level of contamination due to the presence of kaons and hyperons varies with the projectile beam, target nucleus as well as with the incident energy. As nuclear emulsion track detector cannot distinguish between pions and other mesons or hyperons, one possible source of systematic errors is the presence of such contaminations among the pions. Estimation of percentage of contaminations among the pions in nuclear emulsion at the specific energy can be done by applying any event generator [48-49]. UrQMD model can serve this purpose satisfactorily. On UrQMD simulation along with pions K-mesons, $\phi$-mesons, $\Lambda$-hyperons and $\eta$ mesons are also produced. All these mesons can easily be distinguished in terms of their specific codes available from the UrQMD output. Applying the UrQMD model it has been seen that [48-49] for $^{16}$O-AgBr, $^{28}$Si-AgBr and $^{32}$S-AgBr interactions, above 90% of the produced particles are $\pi$-mesons, (3-5) % of the produced particles are kaons and (1-2) % of the produced particles are $\phi$-mesons, $\Lambda$-hyperons and $\eta$ mesons. In our earlier publication [48-49], we have also calculated expected values of average multiplicity of the pions only for all the interactions if the number of other mesons and hyperons could have been evaluated experimentally. We have seen from [48-49] that if we can distinguish between the pions and other mesons as well as $\Lambda$-hyperons, the average multiplicity of only the pions would have been $16.68\pm.23$, $21.58\pm.27$ and $25.54\pm.24$ for $^{16}$O-AgBr, $^{28}$Si-AgBr and $^{32}$S-AgBr interactions respectively [48-49]. The contribution to the systematic errors due to the presence of other mesons and hyperons with the pions in the shower particles has been calculated to be ($7.60\pm.05$ to $8.91\pm.07$) % for these interactions. The total contribution of systematic errors in our analysis does not exceed 10% [49].

## Conclusions

The study of event-by-event maximum pseudo-rapidity gap fluctuations in terms of the scaled variance $\omega$ has been carried out for $^{16}$O-AgBr, $^{28}$Si-AgBr, $^{32}$S-AgBr interactions at an incident momentum of 4.5 AGeV/c applying the multiplicity cut ($N_s > 10$). For all the interactions event-by-event maximum pseudo-rapidity gap fluctuations are found to be greater than zero indicating the presence of strong correlations in the multiparticle production process. The event by event fluctuations of maximum pseudo-rapidity gap are found to decrease with the increase of average multiplicity of the interaction. Experimental analysis has been compared with the results obtained from the analysis of events simulated by generating random numbers (MC-RAND events) and also by Ultra Relativistic Quantum Molecular Dynamics (UrQMD) model. The significant difference of the obtained results of the experimental analysis from the Monte Carlo (MC-RAND) predictions is quite interesting in this analysis. This difference establishes that the observed correlations among the produced shower particles are not merely of statistical origin. Rather the correlations may be interpreted as a consequence of dynamical fluctuation present during the production of the shower particles. UrQMD predicted values of event by event maximum rapidity gap fluctuations decreases with the increase of average multiplicity of the interactions. However, UrQMD model predicts weaker fluctuations in comparison to the experimental analysis.

## References

[1]. S. Jeon *et al.*, Phys. Rev. C **73**, (2006) 014905.

[2]. M. Asakawa *et al.*, Phys. Rev. Lett. **85**, (2000) 2072.

[3]. L.F. Babichev, A.N. Khmialeuski, Study the signals of QGP using Monte Carlo for Pb+Pb for Energy from 500 to 2000 GeV, in Foundations & Advances in Nonlinear Science:Proceedings of 15th International Conference-School,September 20-23, 2010, Minsk, edited by V.I. Kuvshinov,G.G. Krylov (Education and Upbringing, 2010). ISBN 978-985-476-883-0.

[4]. M. Weber for the ALICE collaboration, J. Phys.: Conf.Ser. **389**, (2012) 012036.

[5]. E.V. Shuryak, Phys. Lett. B **423**, (1998) 9.

[6]. A. Bialas and V. Koch, Phys. Lett. B **456**, (1999). 1

[7]. H. Appelsh¨auser et al (NA49 Collaboration), Phys. Lett. B **459**, (1999) 679.

[8]. G. Baym, H. Heiselberg, Phys. Lett. B **469**, (1999) 7.

[9]. G. Danilov, E. Shuryak, Arxiv: nucl-th /9908027v1 6$^{th}$ Aug 1999

[10]    H. Heiselberg  Phys.  Rep. **351**(2001)161

[11]    M.M. Agarwal et al  Phys. Rev. C **65**(2002)054912

[12]    B. Muller and A. Schafer  Phys Rev D **85**(2012) 114030

[13]    S. Jeon and V. Koch Phys. Rev. Lett**. 85** (2000)2076

[14]    B. Abelev et al (ALICE Collaboration) Phys. Rev. Lett. **110**(2013)152301

[15]    F. Karsch Jour. of Phy G **38**(2011) 124908

[16]    V.V Begun et al Phys. Rev. C  **70**(2004)034901

V. V. Begun et al Phys. Rev. C **70**(2004) 034901

V. V. Begun, et al  Phys. Rev C **71**(2005)054904

V. V. Begun et al   Phys. Rev C **76** (2007) 024902

[17]    E.A.De Wolf , I.M.Dremin and  W.Kittel  Phys. Rep. **270**  (1996)48

[18]   D.Ghosh, A. Deb, S. Bhattacharyya and U. Datta Pramana **77**(2011)297

[19]   A. Bialas and R. Peschanski, Nucl. Phys. B**308** (1986)857

       Z. Cao and R. C. Hwa, Phys. Rev. Lett. **75**, (1995)1268;

       Z. Cao and R. C. Hwa, Phys. Rev. D**56** (1997)7361;

       R. C. Hwa and Q. H.Zhang, Phys. Rev. D **62** (2000) 014003.

[20]   R. C. Hwa and Q. H. Zhang, nucl-th/0104019 4 April 2001

       R. C. Hwa and Q. H. Zhang Phys. Rev. C **62** (2000) 054902;

       R. C.Hwa and C. B. Yang, hep-ph/0104216. 25 Feb 2002

[21]   L. Stodolsky, Phys. Rev. Lett. **75** (1995) 1044.

[22]   T. A. Trainor, hep-ph/0001148.17 Jan 2000

[23]   M. Bleicher, et al, Nucl. Phys. A **638** (1998)391c .

[24]   E. Shuryak, Phys. Lett. B **423** (1998)9

[25]   M. Stephanov, K. Rajagopal, and E. Shuryak, Phys. Rev. D **60** (1999) 114028.

[26]   M. Gazdzicki and S. Mrowczynski, Z. Phys. C **54** (1992) 127.

[27]   M. Gazdzicki, Eur. Phys. J. C **8** (1999) 131.

[28]   S. Mrowczynski, Phys. Lett. B **459** (1999)13

[29]   S. Mrowczynski, Phys. Lett. B **439** (1998) 6.

[30]   F. Liu et. al, Euro. Phys. J. C **8** (1999).  649

[31]   S. A. Voloshin, V. Koch and H. G. Ritter   Phy. Rev C **60**(1999) 024901

[32]   M. Bleicher et. al. Phys. Lett. B **435** (1998) 9

[33]   M. Gazdzicki, A. Leonidov and G. Roland, Eur. Phys. J. C **6** (1999) 365.

[34]   S. Mrowczynski, Phys. Lett. B **465** (1999) 8

[35]   G. Baym and H. Heiselberg, Phys. Lett. B **469**  (1999)7 .

[36] M.G. Poghosyan (for the ALICE Collaboration) Jour. of Phy G 38(2011)1240044 Quark-Matter 2011—Proceedings of the 22[nd] International conference on Ultra-relativistic Nucleus-Nucleus Collisions (Annecy, France 23-28 May 2011)

[37] I G Altsybeev (for the ALICE Collaboration)IOp Conf Series Jour of Phys Conf Series 798(2017)012056

[38] M.R. Atayan et al Phy Lett. B 558(2003)29

[39] R.C Hwa and Q.H. Zhang Phy Rev D 62(2000)014003

[40] S. Ahmad et al  Acta Phy Pol B 35(2003)809

[41] D. Ghosh et al Phy Rev C 68(2003)027901

[42] S. Bhattacharyya, M.Haiduc, A.T. Neagu and E. Firu   J. Phys. G: Nucl. Part. Phys. **40** (2013) 025105

[43] C.F.  Powell, P.H. Fowler and D.H. Perkins  "The study of elementary particles by photographic method"  (Oxford, Pergamon,1959) page 450-464 and references therein

[44] S. Bhattacharyya, M.Haiduc, A.T. Neagu and E. Firu   Euro. Phy Jour A 52(2016)301

 [45]    M. Bleicher et al   Jour. of  Phys. G. **25** (1999) 1859

 [46]    S. A. Bass et al    Prog. Part. Nucl. Phys. 41 (1998) 255

[47] S. Bhattacharyya, M.Haiduc, A.T. Neagu and E. Firu    Euro. Phys. Jour. Plus 132(2017)229

 [48]    S. Bhattacharyya, M.Haiduc, A.T. Neagu and E. Firu  International Journal of Modern Physics E   Vol. 23, (2014) 1450012

 [49]    S. Bhattacharyya, M.Haiduc, A.T. Neagu and E. Firu  Jour. of Phys G **41**(2014)075106

**Table 1**

| Interactions at 4.5 AGeV/c | Average Multiplicity of shower particles for the full sample of events | |
|---|---|---|
| | Experimental | UrQMD |
| $^{16}$O-AgBr | 18.05 ± 0.22 | 17.79 ± 0.21 |
| $^{28}$Si- AgBr | 23.62 ± 0.21 | 27.55 ± 0.22 |
| $^{32}$S- AgBr | 28.04 ± 0.14 | 30.84 ± 0.17 |

Table 1 represents the average multiplicities of the shower particles for all the interactions in case of the experimental and the UrQMD data for the full sample of events.

**Table 2**

| Average Multiplicity of shower particles for the events having $N_s > 10$ | | | |
|---|---|---|---|
| Interactions | Experimental | MC-RAND | UrQMD |
| $^{16}$O-AgBr | 21.59 ± 0.22 | 21.59 ± 0.25 | 18.96 ± 0.21 |
| $^{28}$Si- AgBr | 27.44 ± 0.21 | 27.44 ± 0.27 | 27.71 ± 0.22 |
| $^{32}$S- AgBr | 31.32 ± 0.24 | 31.32 ± 0.28 | 30.90 ± 0.17 |

Table 2 represents the average shower particle multiplicities for $^{16}$O-AgBr, $^{28}$Si-AgBr and $^{32}$S-AgBr interactions at 4.5AGeV/c in case of experimental, MC-RAND and UrQMD data after applying the multiplicity cut ($N_s > 10$).

Table 3

| Interactions at 4.5 AGeV/c | Calculated values of $\omega$, the measure of event by event maximum pseudo-rapidity gap fluctuation in high energy nucleus-nucleus interactions | | |
|---|---|---|---|
| | Experimental | MC-RAND | UrQMD |
| $^{16}$O-AgBr | 0.284±0.004 | 0.060±0.001 | 0.253±0.005 |
| $^{28}$Si-AgBr | 0.254±0.005 | 0.041±0.002 | 0.212±0.006 |
| $^{32}$S-AgBr | 0.225±0.009 | 0.083±0.003 | 0.185±0.007 |

Table 3 represents the values of event by event maximum pseudo-rapidity gap fluctuation in high energy nucleus-nucleus insteractions.

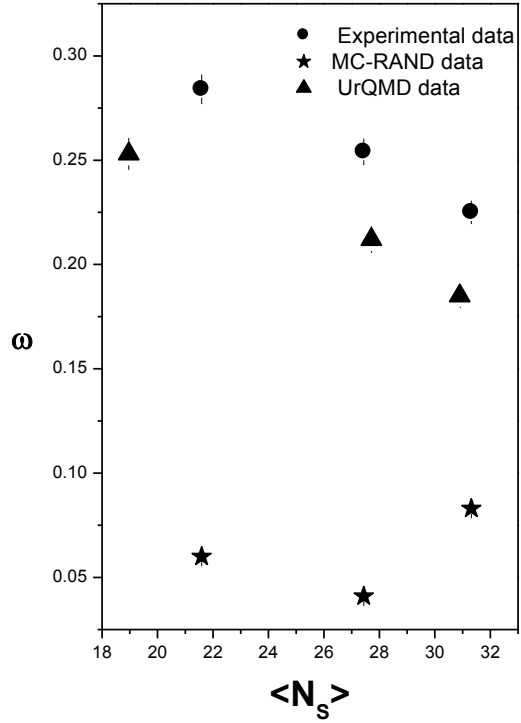

Figure 1: Dependence of the values of event by event maximum pseudo-rapidity gap fluctuation with the average multiplicity of the interactions for the events having $N_s > 10$.

**Acknowledgement**

The authors are grateful to Prof. Pavel Zarubin, JINR, Dubna, Russia for providing them the required emulsion data. Dr. Bhattacharyya also acknowledges Prof. Dipak Ghosh, Department of Physics, Jadavpur University and Prof. Argha Deb Department of Physics, Jadavpur University, for their inspiration in the preparation of this manuscript.